\documentclass[%
 aip,
 amsmath,amssymb,
reprint,%
]{revtex4-1}

\usepackage{graphicx}
\usepackage{dcolumn}
\usepackage{bm}

\usepackage[utf8]{inputenc}
\usepackage[T1]{fontenc}
\usepackage{mathptmx}
\usepackage{xcolor}

\begin{document}

\preprint{AIP/123-QED}

\title{Data-driven subgrid-scale modeling of forced Burgers turbulence using deep learning with generalization to higher Reynolds numbers via transfer learning}

\author{Adam Subel}
\affiliation{
 Department of Mechanical Engineering, Rice University, Houston, TX, USA}

\author{Ashesh Chattopadhyay}%
\affiliation{
 Department of Mechanical Engineering, Rice University, Houston, TX, USA}

\author{Yifei Guan}
\affiliation{
 Department of Mechanical Engineering, Rice University, Houston, TX, USA}

\author{Pedram Hassanzadeh}

\email{pedram@rice.edu}

\affiliation{
 Department of Mechanical Engineering, Rice University, Houston, TX, USA}
\affiliation{
 Department of Earth, Environmental and Planetary Sciences, Rice University, Houston, TX, USA}%


\begin{abstract}
Developing data-driven subgrid-scale (SGS) models for large eddy simulations (LES) has received substantial attention recently. Despite some success, particularly in \textit{a priori} (offline) tests, challenges have been identified that include numerical instabilities in \textit{a posteriori} (online) tests and generalization (i.e., extrapolation) of trained data-driven SGS models, for example to higher Reynolds numbers. Here, using the stochastically forced Burgers turbulence as the test-bed, we show that deep neural networks trained using properly pre-conditioned (augmented) data yield stable and accurate \textit{a posteriori} LES models. Furthermore, we show that transfer learning enables accurate/stable generalization to a flow with $10 \times$ higher Reynolds number.    
\end{abstract}

\maketitle

Due to their high computational cost, the direct numerical simulation (DNS) of turbulent flows will remain out of reach for many real-world applications in the foreseeable future. As a result, the need for parameterization of subgrid-scale (SGS) processes in coarse-resolution models such as large eddy simulation (LES) continues in various areas of science and engineering \citep{pope2001turbulent,sagaut2013multiscale}. In recent years, there has been substantial interest in applications of deep learning for data-driven modeling of turbulent flows \citep{ling2016reynolds,kutz2017deep,brunton2020machine,pathak2018model,wu2020enforcing,mohan2020spatio,chattopadhyay2020ESN,chattopadhyay2020analog,raissi2020hidden,eivazi2020deep,pandey2020perspective}, including for developing data-driven SGS parameterization (DDP) models \citep{pan2018data,duraisamy2019turbulence,xie2019artificialB,maulik2019subgrid2,beck2019deep,zhou2019subgrid,boltonapplications,pawar2020priori,xie2020modeling,chattopadhyay2020data,frezat2020physical,zanna2020data,kurz2020machine,pawar2020interface}. In many of these studies, the goal is to learn the relationship between the filtered variables and SGS terms in high-fidelity data (e.g., DNS data), and use this DDP model in LES. \textit{A priori} tests in some of these studies \citep{maulik2019subgrid2,beck2019deep,zhou2019subgrid,zanna2020data} have shown that such a non-parametric approach can yield DDP models that capture important physical processes (e.g., energy backscatter\citep{piomelli1991subgrid,hewitt2020resolving}) beyond the simple diffusion process that is represented in canonical physics-based SGS models such as Smagorinsky and dynamic Smagorinsky (DSMAG) \cite{smagorinsky1963general,germano1991dynamic,meneveau1997dynamic}. However, these studies have also reported that \textit{a posteriori} (i.e., online) LES tests, in which the DDP model is coupled to a coarse-resolution Navier-Stokes solver, show numerical instabilities or lead to physically unrealistic flows \citep{maulik2019subgrid2,beck2019deep,zhou2019subgrid,kurz2020machine,zanna2020data}. As a remedy, often \textit{ad-hoc} post-processing steps of the DDP models' outputs are introduced, e.g., to remove backscattering or to attenuate the SGS feedback into the numerical solver. Usually, such post-processing steps substantially take away the advantages gained from using deep learning. As a result, numerical instabilities remain a major obstacle to broadening the applications of LES with DDP models.

Another major concern with DDP models is their (in)ability to accurately generalize beyond the flow they are trained for, particularly to flows that have higher Reynolds numbers ($Re$). However, such extrapolations are known to be challenging for neural networks\citep{chattopadhyay2020data,krueger2020out}. Some degree of generalization is essential for building robust and trustworthy LES models with DDP. Furthermore, given that high-fidelity data from often-expensive simulations (e.g., DNS) are needed to train DDP models, some capability to extrapolate to higher $Re$ makes such DDP models much more practically useful.

\begin{figure}[bt]
\includegraphics[width = \linewidth]{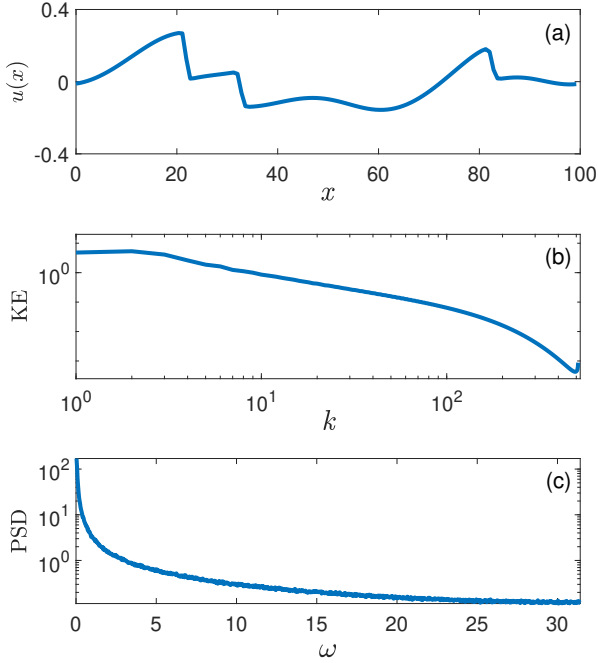}
\caption{\label{fig:DNS_Plots} A sample profile and statistics of the stochastically forced Burgers turbulence (from DNS data at $Re=Re_c$). (a) $u$ showing three distinct shocks; (b) The KE spectrum showing the inertial range; (c) PSD, as a function of frequency $\omega$, showing chaotic behavior.} 
\end{figure}

In this paper, with a particular focus on the issues of stability and generalization, we use a deep artificial neural network (ANN) to develop a DDP model for stochastically forced Burgers turbulence. The forced Burgers equation is\cite{Dolaptchiev2013}   
\begin{equation}
    \frac{\partial u}{\partial t} +\frac{1}{2}\frac{\partial \left(uu\right)}{\partial x} = \nu\frac{\partial^2 u}{\partial x^2} +F,
    \label{eq:Burger_DNS}
\end{equation}
where $u$ is velocity, $\nu=1/Re$, and $F$ is a stochastic forcing (defined later). The domain is periodic with length $L$. Despite being one-dimensional, the presence of strongly nonlinear local regions in the form of shocks, often multiple shocks (Fig.~\ref{fig:DNS_Plots}(a)), makes Burgers turbulence a complex and challenging system, which has been used as the test-bed in various SGS and reduced-order modeling studies\cite{Girimaji1995,Das2002,M.D.1980,Dolaptchiev2013,LaBryer2015,Maulik2018a,alcala2020subgrid}. $F(x,t)$ is defined as\cite{Dolaptchiev2013} 
\begin{equation}
    F = \sum_{k=1}^{3}{\frac{\alpha_{k}A}{\sqrt{20k\Delta t}}\cos\left(2\pi \left(\frac{k x}{L}+\Phi_k \right) \right)},
\label{eq:Forcing}
\end{equation}   
where $k$, $\Delta t$, and $A$ are the wavenumber, time step, and forcing amplitude, respectively. $\Phi_k$ and $\alpha_k$ are a random phase and scaling factor. To develop the LES model, we spatially filter Eq.~(\ref{eq:Burger_DNS}) to obtain
\begin{equation}
   \frac{\partial \bar{u}}{\partial t} +  \frac{1}{2}\frac{\partial\left( \bar{u}\bar{u}\right)}{\partial x} = \nu\frac{\partial^2 \bar{u}}{\partial x^2}+\overline{F}+\Pi,
\label{eq:LES_eq}
\end{equation}
with SGS term
\begin{equation}
    \Pi = -\frac{1}{2}\frac{\partial}{\partial x}\left(\overline{u u} - \overline{\bar{u}\bar{u}}\right).
\label{eq:Pi_Term}
\end{equation}
Here, we use a box filter\citep{pope2001turbulent}. Overbars indicate filtered (and coarse-grained to LES resolution) variables. Note that the difference between $F$ and $\overline{F}$ is negligible. Our aim is to train an ANN to learn $\Pi$ as a function of $\overline{u}$ in the DNS data, and then use this DDP model as a closure in (\ref{eq:LES_eq}).    

We define a setup, referred to as ``control'' and indicated with subscripts ``$c$'', with the following parameters (identical to those used in Dolaptchiev {\it et al}.\citep{Dolaptchiev2013}): $L=100$, $\nu=0.02$, and $A=\sqrt{2}/100$. $\Phi_k$ and $\alpha_k$ are drawn randomly from $\mathcal{N}(0,1)$ every $20\Delta t$ to update $F$. To obtain the DNS data, which are treated as the ``truth'', Eq.~(\ref{eq:Burger_DNS}) is integrated using a pseudo-spectral solver with $1024$ Fourier modes and time step $\Delta t=0.01$. Figure~\ref{fig:DNS_Plots} shows a sample profile of $u(x)$, and the kinetic energy (KE) spectrum and power spectral density (PSD) of the flow. To perform LES, Eq.~(\ref{eq:LES_eq}) with the DDP model of $\Pi(\overline{u})$ is integrated using the same pseudo-spectral solver but with $128$ Fourier modes and time step $20\Delta t$. The schematic of LES with DDP is shown in Fig.~\ref{fig:DDP_Ils}(a). Details of the ANN and the training data/procedure are presented below.

We use a multilayer perceptron ANN\citep{goodfellow2016deep} to develop the DDP model. This ANN is unidirectional (information only passes in one direction from input to output) and is fully connected between the layers. The ANN is trained, i.e., all learnable parameters of the network (weights and biases, collectively represented by $\theta$) are computed, by minimizing the mean-square-error $MSE=\sum_{i=1}^{M}\|ANN(\tilde{\overline{u}}_i;\theta) -\tilde{\Pi}_i \|_2^2/M$. Here, $M$ is the number of training samples, $\| \cdot \|_2$ is the $L_2$ norm, $\overline{u}$ and $\Pi$ are calculated from DNS data, and $\tilde{\cdot}$ indicates pre-conditioned (augmented) training data (discussed shortly). The best network architecture, found based on extensive trial and error using $MSE$, consists of an input layer, $6$ hidden layers with $250$ nodes each, and a linear output layer. On all but the final layer, the swish activation function\cite{ramachandran2017searching} is used. 

\begin{figure}[bt]
\includegraphics[width = \linewidth]{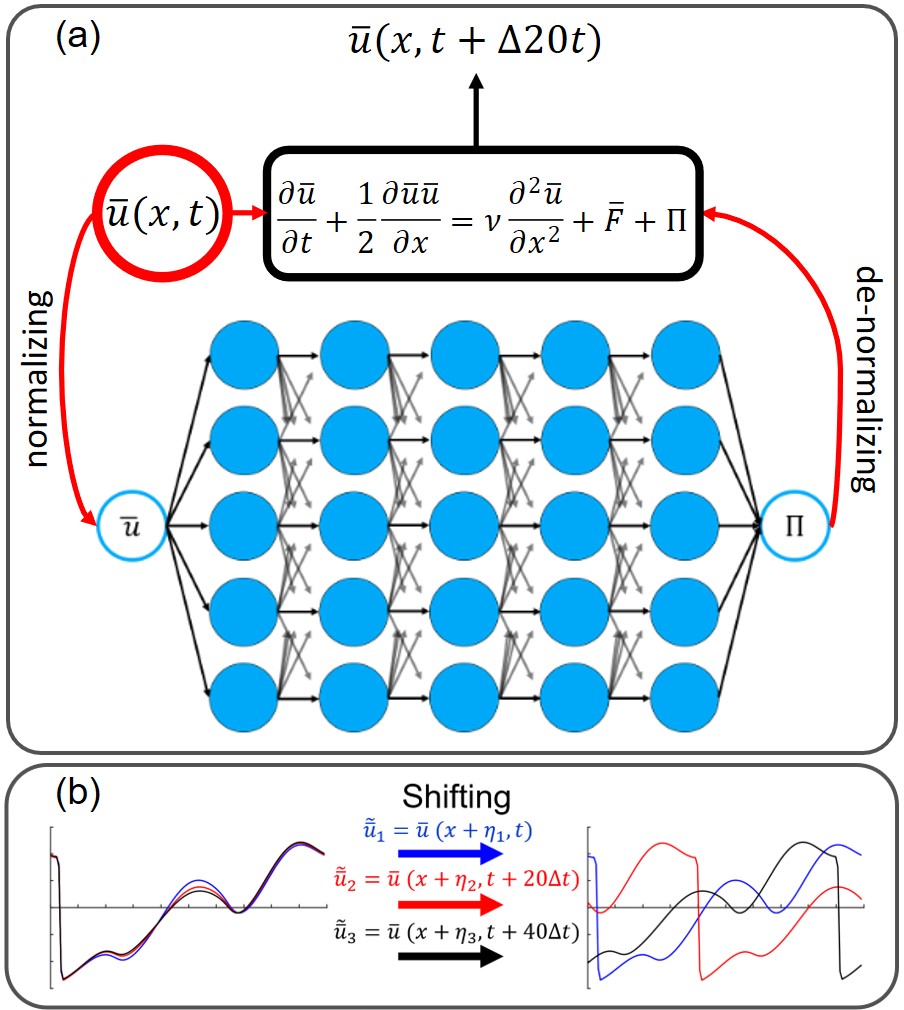}
\caption{\label{fig:DDP_Ils} (a) The schematic of the LES with DDP model. With normalized $\overline{u}(x,t)$ as input, the trained ANN predicts $\Pi$, which is then de-normalized and used in Eq.~(\ref{eq:LES_eq}) to compute $\overline{u}(x,t+20\Delta t)$, and the cycle continues. (b) The pre-conditioning step to augment the training data by adding random shifts in $x$ to produce spatially diverse samples from a relatively small DNS dataset.
}
\end{figure}

Our first attempts to train the DDP model with even a relatively large training set, $M=O(10^5)$, resulted in inaccurate $\Pi$ terms in \textit{a priori} tests and unstable LES with DDP in \textit{a posteriori} tests. Further analysis showed that the problem is due to the fact that the SGS dynamics and thus the $\Pi$ terms in Burgers turbulence are highly localized around the shocks\cite{M.D.1980}, which as explained below, leads to overfitting, i.e., poor generalization of ANN (at the same $Re$) beyond the training set. Shocks are persistent and can remain fairly stationary for many time steps, which can lead to small or near-zero $\Pi$ terms in some regions of the domain that do not experience shocks throughout the entire training set. The ANN trained on such a dataset will predict $\Pi \approx 0$ in those regions no matter what the inputted $\overline{u}$ is during (\textit{a priori} or \textit{a posteriori}) tests. Note that by design, the flow during training could be very different, in terms of the location of shocks and their evolution, from the flow during testing (though the training and testing sets have the same $Re$, the latter is chosen from an independent DNS run or from a time window far from the time window of the training set). Of course, this overfitting problem can be resolved by using a much larger training set that contains a sufficient number of samples of shocks waves occurring in all regions; however, such large training sets are often unavailable. Here, we propose a simple strategy, based on pre-conditioning the training samples, to overcome this problem without the need for a larger dataset.

As shown in Fig.~\ref{fig:DDP_Ils}(b), a random shift $\eta$, drawn from the uniform distribution $\mathcal{U}(0,L)$, is added to $x$ for each input-output pair $(\overline{u},\Pi)$   
\begin{equation}
    \label{eq:shift}
    \Tilde{\overline{u}}(x,t) = \overline{u}(x - \eta,t)
     \; \; \mathrm{and} \; \;   \Tilde{\Pi}(x,t)= \Pi(x - \eta,t).
\end{equation}
The periodicity in $x$ is used when $x - \eta < 0$. It should be noted that this type of artificially enhancing the richness of information inside the training set is commonly used in the machine learning community and is called data augmentation\cite{tanner1987calculation}. For example, in processing of natural images, data augmentation generally involves artificially enhancing the training set by rotating, mirroring, or cropping images. Here, we have exploited the periodicity of $x$ to introduce a physically meaningful augmentation, which allows us to enrich the information of the localized flow and SGS terms around shock waves in the training set without the need for a longer DNS dataset. Finally, as is common practice in machine learning, the input $\tilde{\overline{u}}$ and output $\tilde{\Pi}$ samples are separately normalized (through removing the mean and  dividing by the standard deviation).

The pre-conditioned input-output pairs $(\tilde{\overline{u}},\tilde{\Pi})$ are used to train the ANN. As shown next, the DDP model with an ANN trained using augmented data leads to accurate $\Pi$ terms in \textit{a priori} tests and stable and accurate LES models in \textit{a posteriori} tests without the need for any post-processing of the trained ANN or its output (with the exception of de-normalizing the predicted $\Pi$; see Fig.~\ref{fig:DDP_Ils}(a)). We have used $M=5\times 10^5$ samples for training and another (independent) $5\times 10^4$ samples for validation from a DNS run at $Re=Re_c$. For testing, we have used data from the same run but $5\times 10^4 \Delta t$ separated from the training/validation sets as well as data from two other independent DNS runs at $Re=Re_c$.      





\begin{figure}[bt]
\includegraphics[width = 1\linewidth]{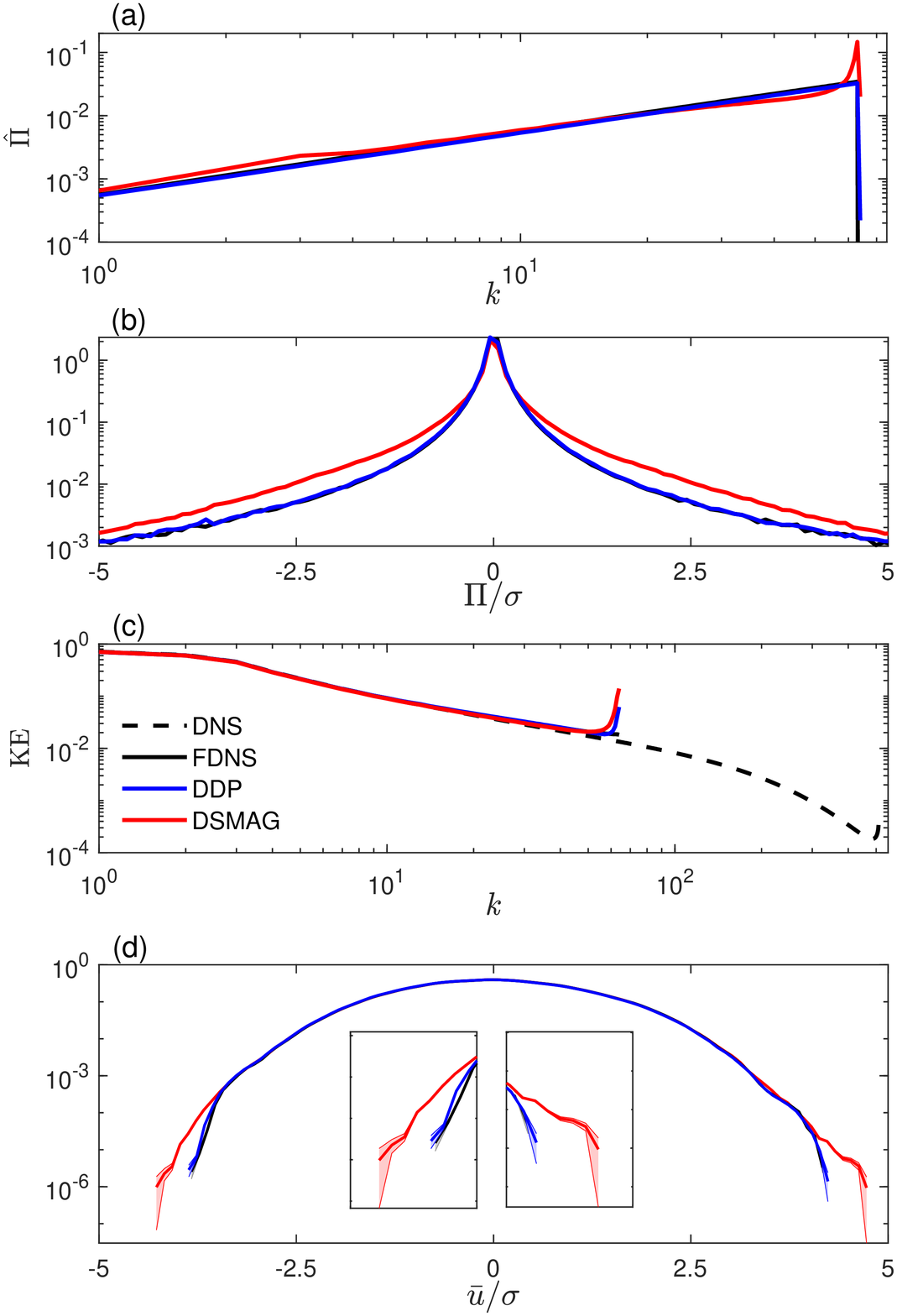}
\caption{\label{fig:DDP_small} Statistics of the resolved flow $\overline{u}$ and SGS term $\Pi$ calculated using results from \textit{a posteriori} tests at $Re=Re_c$. The training and testing data are both at $Re=Re_c$. (a) spectrum of $\Pi$, denoted as $\hat{\Pi}(k)$. The spectrum for FDNS agrees with those reported in previous studies of Burgers turbulence\cite{Das2002}. (b) PDF of $\Pi$. (c) spectrum of KE. The curl up in KE around the maximum resolved $k$ of LES is a common feature of spectral LES solvers applied to Burgers turbulence\cite{LaBryer2015,Maulik2018a,Bayona2018}. In (a)-(c), each curve is produced using $3\times 10^5$ sequential samples that are $20\Delta t$ apart. (d) PDF of $\bar{u}$ computed using a kernel estimator\cite{wilcox2010fundamentals}. Inset panels in (d) show the zoomed-in left and right tails. Shading shows uncertainty as $\pm 1$ standard deviation obtained from bootstrapping $3$ independent LES or DNS runs that are combined (each providing $3\times 10^5$ samples as before) . In (b) and (d), $\sigma$ is the variable's standard deviation.}
\end{figure}

We examine the performance of the LES with DDP in \textit{a posteriori} (online) tests to assess both accuracy (of the SGS modeling) and stability of the hybrid model. Given that the numerical solution of Eq.~(\ref{eq:LES_eq}) blows up without any SGS modeling (i.e., with $\Pi=0$), we use a conventional SGS scheme, DSMAG\cite{Yanan2016}, as the baseline. Figure~\ref{fig:DDP_small}(a)-(b) shows the spectrum and the probability density function (PDF) of the $\Pi$ terms predicted by DDP and DSMAG compared against those of the filtered DNS (FDNS), which is treated as the truth. Both panels show that the statistics of $\Pi$ predicted by DDP closely follow those of the truth at any $k$ and even at the tails of the PDF. Furthermore, both panels show that DDP outperforms DSMAG in modeling the statistics of the SGS term ($\Pi$). The better performance of DDP is clearly seen at high and low $k$ in (a) and beyond $\pm 1$ standard deviation in (b). Note that the difference between the $\Pi$'s PDFs from FDNS and DSMAG (DDP) is (is not) statistically significant at $95\%$ confidence level based on both Kolmogorov-Smirnov, KS, and Kullback–Leibler divergence, KL, tests\cite{wilcox2010fundamentals}.



To examine the statistics of the resolved flow, Fig.~\ref{fig:DDP_small}(c)-(d) shows the spectrum of KE and the PDF of $\overline{u}$. Both LES with DDP and LES with DMSAG capture the KE spectrum up to near the maximum resolved $k$ ($=64$) although DDP does slightly better and agrees with the FDNS' KE spectrum up to $k \approx 60$ while DSMAG does so up to $k \approx 50$. Furthermore, as shown in panel (d), LES with DDP outperforms LES with DSMAG in capturing the PDF's tails, which correspond to shocks. Note that the differences between the PDFs of DDP, FDNS, and DSMAG are not statistically significant (at $95\%$ confidence level) based on the KS or KL test, but that is because such tests mainly assess similarities in the bulk rather than the tails of the PDFs. A closer visual inspection shows that the difference between the tails of the PDFs from FDNS and DDP (DSMAG) is within (outside) the uncertainty range, indicating that DDP (DSMAG) accurately captures (does not capture) the statistics of the rare events. 

In summary, the DDP model that uses an ANN trained with augmented data (from $Re=Re_c$) leads to a stable LES model in \textit{a posteriori} tests (at $Re=Re_c$) that is more accurate than LES with DSMAG. Next, we examine whether a DDP model trained with augmented data from a given $Re$ can be used for LES of a flow that has higher $Re$.    



\begin{figure}[t]
\includegraphics[width = \linewidth]{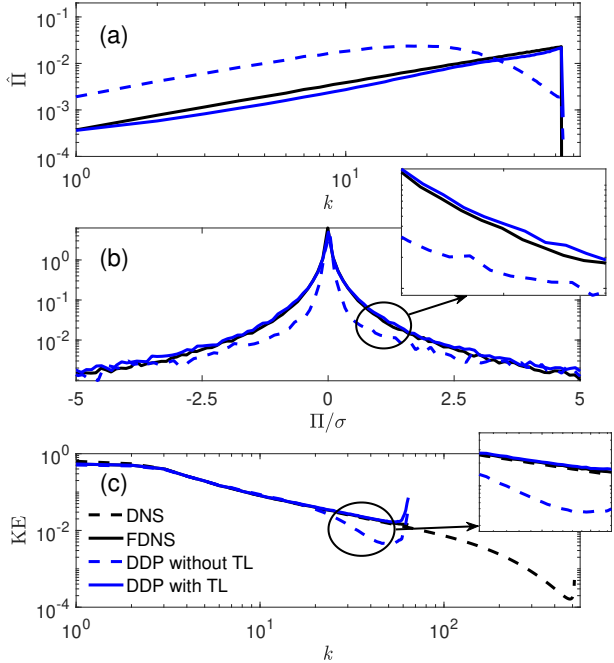}
\caption{\label{fig:Re_10} Statistics of the resolved flow and SGS term calculated using results from \textit{a posteriori} tests at $Re=Re_c$ but with DDP models mainly trained on data from $Re=Re_c/10$. Each curve is produced using $3 \times 10^5$ sequential samples that are $20\Delta t$ apart. The DDP model without transfer learning (TL) uses the ANN trained on $M=5\times10^5$ samples from DNS at $Re=Re_c/10$. The DDP model with TL uses the same ANN but after its last two layers are re-trained with $5\times10^4$ samples from DNS at $Re=Re_c$ (Fig.~\ref{fig:TL_Cartoon}). (a) spectrum of $\Pi$. (b) PDF of $\Pi$. (c) spectrum of KE.}
\end{figure}

Figure~\ref{fig:Re_10} shows the statistics of the resolved flow and of $\Pi$ calculated using results from \textit{a posteriori} tests at $Re=Re_c$ but with a DDP model that uses an ANN trained on data from $Re=Re_c/10$ (see the dashed blue lines). It is clear that this DDP model \textit{does not} generalize as the spectrum and PDF of $\Pi$ and the spectrum of KE all deviate from those of the FDNS. The results are not surprising as it is known that data-driven models often have difficulty with generalization to a different (especially more complex) system. For example, using a multi-scale Lorenz 96 system, we~\cite{chattopadhyay2020data} showed that ANN- and recurrent neural network-based data-driven SGS models do not accurately generalize when the system is forced to become more chaotic. However, we also showed that transfer learning (TL)\cite{yosinski2014transferable} provides an effective way for addressing this challenge, at least for a simple chaotic toy model. Below, we show the effectiveness of TL in making DDP generalizable to higher $Re$ in a turbulent flow.

\begin{figure}[t]
\includegraphics[width = \linewidth]{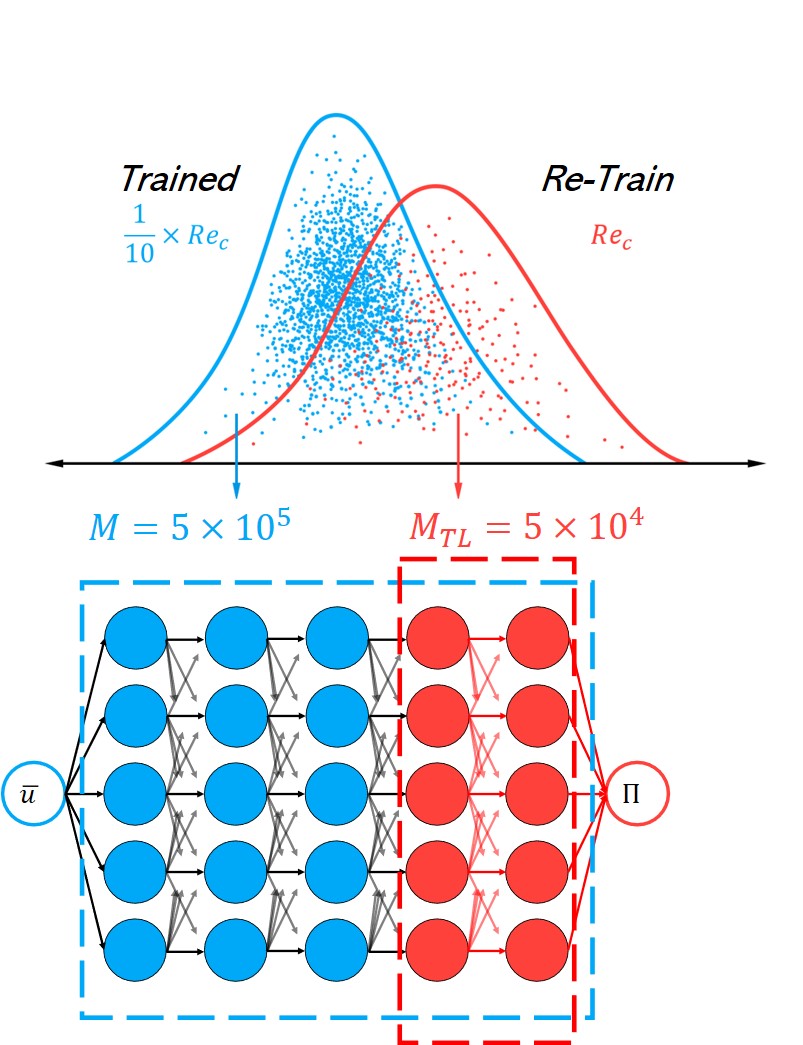}
\caption{\label{fig:TL_Cartoon} Schematic of transfer learning (TL) to develop an accurate DDP model for $Re=Re_c$. Without TL, the ANN in the DDP model is trained, starting with random weights, on {$M=5\times 10^5$} samples from DNS at {$Re=Re_c/10$}. This DDP model \textit{does not} generalize to $Re=Re_c$ (dashed blue lines in Fig.~\ref{fig:Re_10}). Then, TL is applied: the weights in the first three layers (blue) of this ANN are fixed, and the last two layers (red) are re-trained, starting with the previously computed weights, and using only {$M_{TL}=5\times 10^4$} samples from DNS at $Re=Re_c$. The DDP model with TL is accurate and stable in \textit{a posteriori} tests at $Re=Re_c$ (solid blue lines in Fig.~\ref{fig:Re_10}).}
\end{figure}

Figure~\ref{fig:TL_Cartoon} shows the schematic of TL applied to the ANN of a DDP model. In general, the weights of an ANN are randomly initialized and then they are updated through training on $M$ samples from a given data distribution (here, data from a flow with $Re=Re_c/10$). The test in Fig.~\ref{fig:Re_10} showed that this ANN does not accurately work for $Re=Re_c$. The idea of TL is that we re-train this ANN (starting with its current weights rather than random initializations) and update the weights only in the deeper layers using a smaller number of samples (e.g., $M_{TL}=M/10$) from the new data distribution (i.e., the flow with $Re=Re_c$). The underlying idea of TL is that in deep networks, the initial layers learn high-level features, and only the deeper layers learn low-level features that are specific to a particular data distribution\cite{yosinski2014transferable}. Thus, for generalization, we only need to re-train the deeper layers, which can be done using a small amount of data from the new distribution. 

\begin{figure}[t]
\includegraphics[width = \linewidth]{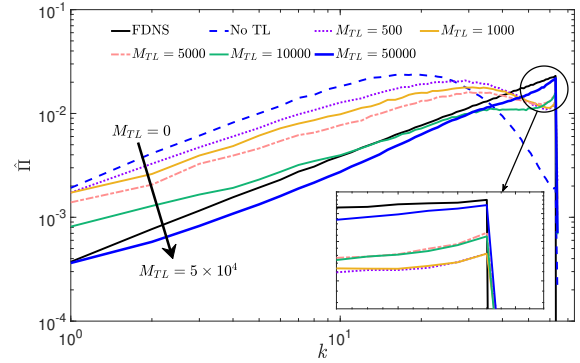}
\caption{\label{fig:Scaling} Spectrum of $\Pi$ in \textit{a posteriori} tests on $Re=Re_c$ as $M_{TL}$ (the number of training samples from $Re_c$ used in TL) is increased. $M_{TL}=0$ correspond to no TL and the original ANN trained on $M=5\times10^5$ samples from $Re=Re_c/10$. Adding $M_{TL}=500$ to $5000$ samples improves the generalization capability of the DDP model to some degree. $M_{TL}=10^4$ ($2\%$ of $M$) leads to substantial improvements although $\hat{\Pi}$ is underestimated at high $k$ while overestimated at low $k$. Increasing $M_{TL}$ to $5\times 10^4$ ($10\%$ of $M$) further improves the generalization capability and $\hat{\Pi}$ that is just slightly underestimated.}
\end{figure}

Figure~\ref{fig:Re_10} shows that the DDP model with TL (solid blue lines) accurately generalizes to the flow with $Re_c$ as the spectrum and PDF of $\Pi$ and spectrum of KE closely match those of FDNS. In fact, the accuracy of the DDP model with TL in Fig.~\ref{fig:Re_10} (which only uses $M_{TL}=5\times10^4$ training samples from $Re_c$), is comparable with the accuracy of the DDP model in Fig.~\ref{fig:DDP_small} (which uses $M=5\times10^5$ training samples from $Re_c$). Finally, Fig.~\ref{fig:Scaling} shows how gradually increasing $M_{TL}$ improves the generalization capability of the DDP model.

In conclusion, we have investigated ANN-based data-driven SGS modeling of Burgers turbulence, with a particular focus on the stability of \textit{a posteriori} LES models and generalization to higher $Re$. We show that developing a DDP model for Burgers turbulence is particularly challenging due to the presence of shocks, which localize the SGS term ($\Pi$), resulting in ANNs that overfit in the absence of a large training set. The overfitting ANNs lead to inaccurate/unstable DDP models. To overcome this challenge, we introduce a pre-conditioning step in which, exploiting periodicity, training samples are randomly shifted, thus enriching and augmenting the training set. The DDP model trained on this augmented dataset leads to stable and accurate \textit{a posteriori} LES models. These results suggest that similar data augmentation strategies that exploit symmetries and other physical properties should be considered in developing DDP models for more complex flows when large training sets are unavailable, not only to improve accuracy but also to improve the stability of \textit{a posteriori} LES runs. 

We have also found the DDP model not to generalize (i.e., extrapolate) to a flow with $10\times$ higher $Re$. However, we show, for the first time to the best of our knowledge, the application of TL to making a DDP model generalizable in a turbulent flow.  Transfer learning enables the development of DDP models for high-$Re$ flows with most of the training data provided by high-fidelity simulations at lower $Re$, which is highly appealing for practical purposes because the computational cost of simulating turbulent flows rapidly increases with $Re$.    

In future work, the application of TL and data augmentation to develop accurate, stable, generalizable DDP models for more complex turbulent flows that are 2D and 3D will be investigated.  




\section*{Data Availability}
The training and validation datasets are openly available in Zenodo at \url{https://doi.org/10.5281/zenodo.4316338}. The DNS and LES solvers, data analysis codes, and machine learning codes are publicly available in GitHub at \url{https://github.com/envfluids/Burgers_DDP_and_TL}.

\begin{acknowledgments}
We thank Romit Maulik, Rambod Mojgani, and Ebrahim Nabizadeh for insightful discussions. This work was supported by an award from the ONR Young Investigator Program, N00014-20-1-2722, and by NSF grant OAC-2005123 (to P.H.). A.C. thanks the Rice University Ken Kennedy Institute for Information Technology for a BP HPC Graduate Fellowship. Computational resources were provided by NSF XSEDE (allocation ATM170020) and by the Rice University Center for Research Computing. 
\end{acknowledgments}



\bibliography{PoF_Burgers_v2}

\end{document}